\documentstyle[preprint,aps]{revtex}
\begin{document}
\draft
\title{
Generalized $CP^1$ model from $t_1-t_2-J$ model
	 }
\author{Naoum Karchev\cite{byline}}
\address{
Department of Physics, University of Sofia, 1126 Sofia, Bulgaria
	  }
\maketitle
\begin{abstract}
A long-wavelength, low-frequency effective theory is obtained from $t_1-t_2-J$
model. The action is written in terms of two-component bose spinor fields
($CP^1$ fields) and two spinless Fermi fields. The generalized $CP^1$ model
is invariant under $U(1)$ gauge transformations. The bose fields and
one of the Fermi fields have charge $+1$ while the other Fermi field has charge
$-1$ with respect to these transformations. A simple mean-feild theory
of a gauge-symmerty breaking, based on a four-fermion interaction, is
discussed. An effective theory of frustrated antiferromagnetism is obtained
integrating out the Fermi fields around the mean-fields.

Another option is used to parametrize the long distance fluctuations in
$t_1-t_2-J$ model, with the help of gauge invariant fields. It is argued
that the resulting Fermi quasiparticles of the $t_1-t_2-J$ model have both
charge and spin. The effective action is rewritten in terms of spin
$\frac 12$ Fermi spinor, which has the charge of the holes, and unit vector.
\end{abstract}
{\bf I Introduction}

High-$T_c$ superconductivity has given theorists a strong motivation to
work on correlated electrons. Among the many electronic models that are
being currently studied, the two-dimensional $t-J$ model is the simplest one
which captures the essential physics of strongly correlated electronic
systems. Anderson \cite{1} first applied this model to high $T_c$-oxides.
Zhang, Rice \cite{2} and others \cite{2a} showed that one band $t-J$
model is an effective model describing the physics of the three band
$Cu-O$ model \cite{3}.

The underlying problem is to create an adequate field theory. Usually the
bosonic and fermionic raising and lowering operators are used to realize
spin-fermion algebra. Several approximate techniques have evolved so
far to deal with the $t-J$ model. A representation for the electron
operators acting on states with no double occupancy has been proposed
\cite{4} in terms of spin fermion operators and spinless boson which keeps
track of empty sites. Mean-field theory of high-$T_c$ superconductivity
based on this representation has been developed \cite{5}. The slave-boson
technique is widely applied to the study of various properties of the
model \cite{6}, and to a large range of problems: Hubbard model \cite{6a},
Kondo lattice model \cite{4,6b}, the Anderson Hamiltonian\cite{6c}.

Mean-field theory based on the alternative Schwinger-bosons slave-fermion
representation has been worked out \cite{7}. Similar mean-field approach
has been used to investigate the phase diagram of the $t_1-t_2-J$ model
\cite{8}.

It is important to stress upon the fact that these theories start with
one and the same
Hamiltonian, and use equivalent representations of the spin-fermion
algebra. But these representations allow different appropriate methods of
approximate calculations which may arrive at completely different
description of the properties of the model. The mean-feild approximations
are self-consistent, but it is difficult to judge how close to the true
properties of the model the results are.

Numerical calculations have been done using a large number of techniques
\cite{9}. The results of these calculations can contribute to the
acceptance or rejection of mean-field based theories, and can also
indicate directions in which new aprroaches should be developed.

The reduction of the three-band model to the one-band $t-J$ model
is still controversial. Many authors have argued that the resulting
quasiparticles of the three-band model have both charge and spin.
The effective spin-fermion model is characterized by
a Kondo like coupling of the $O$ holes to localized $Cu$ spins and a
Heisenberg antiferromagnetic interaction among $Cu$ spins \cite{10}.
The long-wavelength limit of the model has been written in terms of fermionic
spinors and a unit vector \cite{11}. The dynamic of the order parameter of the
spin background is given by the $0(3)$ nonlinear $\sigma$-model, and the
interaction of the mobile holes with the order parameter is a
current-current type of interaction.

The three-band $Cu-O$ model contains strong interactions, and the perturbative
calculation of bubble and ladder diagrams are questionable. An analytical
method, that seems to be able to handle strong correlations has been
proposed \cite{12}. The photoemission and inverse photoemission spectra of
holes calculated by means of the projection technique reproduce the numerical
results.

It is widely accepted that the undoped oxides superconductors can be modeled
rather well by a nearest-neighbor $s=\frac 12$ antiferromagnetic Heisenberg
Hamiltonian on a square lattice.
It has been argued that the long-wavelength, low-temperature behavior of
the model can be described by a quantum nonlinear $\sigma$ model \cite{13}.
The low-temperature behavior of the correlation length and the static and
dynamic spin-correlation functions have been calculated using the
renormalization group method \cite{14}. The results are in good agreement
with the experimental data.

The present work is motivated
by the successfull application of the quantum mechanical non-linear
$\sigma$ model to high-$T_c$ oxides. A long-wavelength, low-frequency
effective theory is obtained from $t_1-t_2-J$
model. The action is written in terms of two-component bose spinor fields
($CP^1$ fields) and two spinless Fermi fields. The generalized $CP^1$ model
is invariant under $U(1)$ gauge transformation. The bose fields and
one of the Fermi fields have charge $+1$ while the other Fermi field has charge
$-1$ with respect to these transformations. A simple mean-feild theory
of the gauge-symmerty breaking, based on the four-fermion interaction, is
discussed. An effective theory of frustrated antiferromagnetism is obtained
integrating out the Fermi fields around the mean-fields.

Another option is used to parametrize the long distance fluctuations in
$t_1-t_2-J$ model, with the help of gauge invariant fields. It is argued
that the resulting Fermi quasiparticles of the $t_1-t_2-J$ model have both
charge and spin. The effective action is rewritten in terms of spin
$\frac 12$ Fermi spinor, which has the charge of the holes, and unit vector.

The paper is organized as follows: Sec $II$ is devoted to the derivation of the
generalized $CP^1$ model from $t_1-t_2-J$ one. In Sec $III$ a dynamical
breakdown of the gauge symmetry is discussed. The effective action is
rewritten in terms of gauge invariant fields in sec $IV$. The charge and
the spin of the Fermi quasiparticles are discussed. Finally, I comment on
the relations to the other effective models.
\ \\

{\bf II Generalized $CP^1$ model}

The $t_1-t_2-J$ model is defined by the Hamiltonian
\FL
\begin{eqnarray}
h\,= & t_1 & \sum\limits_{<i,j>}\left[c^+_{i\sigma}c_{j\sigma}+h.c.\right]\,
+\,t_2\sum\limits_{<<i,j>>}\left[c^+_{i\sigma}c_{j\sigma}+h.c.\right]
\nonumber \\
& + & \,J\sum\limits_{<i,j>}\left({\vec S_i}\cdot{\vec S_j}\,-\,\frac 14
n_i\,n_j \right)\,-\,\mu\,\sum\limits_i\,n_i.
\end{eqnarray}

The electron operators $c_{i\sigma}\left(c^+_{i\sigma}\right)$, the spin
operators $\vec S_i$, and the number operator $n_i$ act on a restricted
Hilbert space where the doubly-ocupied state is excluded. The sums are
over all sites of a two-dimensional square lattice, $<i,j>$ denotes the sum
over the nearest neighbors, and $<<i,j>>$ denotes the sum over the next to
nearest neighbors.

Let us represent the eight operators by means of Schwinger bosons
$\varphi_{i\sigma}\,\left(\bar\varphi_{i\sigma}\right)$, $\sigma\,=\,1,2$ and
slave-fermions $\psi_i\,\left(\bar\psi_i\right)$
\FL
\begin{eqnarray}
c_{i\sigma}\,=\,\bar \psi_i\,\varphi_{i\sigma},\hspace{2cm}
& \vec S_i\,=\,\frac 12\,\bar\varphi_i\vec\sigma\varphi_i,\nonumber \\
\bar c_{i\sigma}\,=\,\psi_i\,\bar\varphi_{i\sigma},\hspace{2cm}
& n_i\,=\,1-\bar\psi_i\,\psi_i,
\end{eqnarray}

where $\vec\sigma$ are Pauli matrices. The finite dimensional space of
representation is a subspace of the Hilbert space of bosons and fermions
defined by the operator consraint
\FL
\begin{equation}
\bar\varphi_{i\sigma}\varphi_{i\sigma}\,+\,\bar\psi_i\psi_i\,=\,1
\end{equation}

The partition function can be written as a path integral over the complex
functions of the Matsubara time $\tau$\,\, $\varphi_{i\sigma}(\tau)\,
\left(\bar\varphi_{i\sigma}(\tau)\right)$ and Grassmann functions
$\psi_i(\tau)\,\left(\bar\psi_i(\tau)\right)$
\FL
\begin{equation}
\cal Z(\beta)\,=\,\int\,d\mu\left(\bar\varphi,\varphi,\bar\psi,\psi\right)
e^{-S}.
\end{equation}

The action is given by the expression
\FL
\begin{equation}
S\,=\,\int\limits^{\beta}_0 d\tau\left[\sum\limits_i\left(\bar\varphi_{i\sigma}
(\tau)
\dot\varphi_{i\sigma}(\tau)\,+\,\bar\psi_i(\tau)\dot\psi_i(\tau)\right)\,+\,
h\left(\bar\varphi,\varphi,\bar\psi,\psi\right)\,\right],
\end{equation}
where $\beta$ is the inverse temperature and the Hamiltonian is obtained from
Eq.(1) replacing the operators with the functions. In terms of Schwinger-bosons
and slave-fermion the theory is $U(1)$ gauge invariant, and the measure
includes $\delta$-functions which enforce the constraint Eq.(3) and the
gauge-fixing condition
\FL
\begin{eqnarray}
d\mu\left(\bar\varphi,\varphi,\bar\psi,\psi\right) & = &
 \prod\limits_{i,\tau,\sigma}\frac {d\bar\varphi_{i\sigma}(\tau)
d\varphi_{i\sigma}(\tau)}{2\pi i}\prod\limits_{i\tau}d\bar\psi_i(\tau)
d\psi_i(\tau)\prod\limits_{i\tau}\delta\left(g.f\right) \nonumber \\
& & \prod\limits_{i\tau}\delta\left(\bar\varphi_{i\sigma}(\tau)\varphi_{i\sigma}(\tau)
\,+\,\bar\psi_i(\tau)\psi_i(\tau)\,-\,1\right).
\end{eqnarray}

I make a change of variables, introducing new bose fields $f_{i\sigma}(\tau)
\left(\bar f_{i\sigma}(\tau)\right)$ \cite{15}
\FL
\begin{eqnarray}
\varphi_{i\sigma}(\tau) & = & f_{i\sigma}(\tau)\sqrt {1\,-\,\bar\psi_i(\tau)
\psi_i(\tau)}\,=\,f_{i\sigma}(\tau)\left(1\,-\,\frac 12\bar\psi_i(\tau)
\psi_i(\tau)\right) \nonumber \\
\bar\varphi_{i\sigma}(\tau) & = & \bar f_{i\sigma}(\tau)
\sqrt {1\,-\,\bar\psi_i(\tau)
\psi_i(\tau)}\,=\,\bar f_{i\sigma}(\tau)\left(1\,-\,\frac 12\bar\psi_i(\tau)
\psi_i(\tau)\right).
\end{eqnarray}
The inverse relations are
\FL
\begin{eqnarray}
f_{i\sigma}(\tau) & = & \varphi_{i\sigma}(\tau)\left(1\,+\,\frac 12\bar\psi_i(\tau)
\psi_i(\tau)\right) \nonumber \\
\bar f_{i\sigma}(\tau) & = & \bar \varphi_{i\sigma}(\tau)\left(1\,+
\,\frac 12\bar\psi_i(\tau)\psi_i(\tau)\right),
\end{eqnarray}
and it is easy to see that the new fields satisfy the constraint
\FL
\begin{equation}
\bar f_{i\sigma}(\tau)f_{i\sigma}(\tau)\,=\,1.
\end{equation}

Inserting (7) into Eqs.(1-6) one obtains for the action
\FL
\begin{equation}
S\,=\,\int\limits^{\beta}_0 d\tau \left\{\sum\limits_i\left[\bar f_{i\sigma}(\tau)
\dot f_{i\sigma}(\tau)\left(1-\bar\psi_i(\tau)\psi_i(\tau)\right)\,+\,
\bar\psi_i(\tau)\dot\psi_i(\tau)\right]\,+\,h\left(\bar f,f,\bar\psi,\psi\right)
\right\}
\end{equation}
where the Hamiltonian is
\FL
\begin{eqnarray}
h\,= & - & t_1\sum\limits_{<i,j>}\left[\bar\psi_i(\tau)\psi_j(\tau)
f_{i\sigma}(\tau)\bar f_{j\sigma}(\tau)\,+\,\bar\psi_j(\tau)\psi_i(\tau)
f_{j\sigma}(\tau)\bar f_{i\sigma}(\tau)\right] \nonumber \\
& - & t_2\sum\limits_{<<i,j>>}\left[\bar\psi_i(\tau)\psi_j(\tau)
f_{i\sigma}(\tau)\bar f_{j\sigma}(\tau)\,+\,\bar\psi_j(\tau)\psi_i(\tau)
f_{j\sigma}(\tau)\bar f_{i\sigma}(\tau)\right] \nonumber \\
\, & - & \frac J4\sum\limits_{<i,j>}\left[1\,-\,\bar f_i(\tau)\vec\sigma
f_i(\tau)\bar f_j(\tau)\vec\sigma f_j(\tau)\right]
\left(1\,-\,\bar\psi_i(\tau)\psi_i(\tau)\right)
\left(1\,-\,\bar\psi_j(\tau)\psi_j(\tau)\right) \nonumber \\
\, & - & \mu\sum\limits_i\left(1\,-\,\bar\psi_i(\tau)\psi_i(\tau)\right).
\end{eqnarray}
The partition function can be written as an integral over the fields
$f_{i\sigma}(\tau),\bar f_{i\sigma}(\tau), \psi_i(\tau), \bar\psi_i(\tau)$
and the measure is given by the equality
\FL
\begin{eqnarray}
d\mu\left(\bar f,f,\bar\psi,\psi\right) & = &
 \prod\limits_{i,\tau,\sigma}\frac {d\bar f_{i\sigma}(\tau)
d f_{i\sigma}(\tau)}{2\pi i}\prod\limits_{i\tau}d\bar\psi_i(\tau)
d\psi_i(\tau)\prod\limits_{i\tau}\delta\left(g.f\right) \nonumber \\
& & \prod\limits_{i\tau}\delta\left(\bar f_{i\sigma}(\tau) f_{i\sigma}(\tau)
\,-\,1\right)\prod\limits_{i\tau} e^{-\bar\psi_i(\tau)\psi_i(\tau)}.
\end{eqnarray}
where the last multiplier just redefines the chemical potential.

I consider two sublattices $A$ and $B$, and impose the gauge-fixing
conditions in the form
\FL
\begin{eqnarray}
& & arg f_{i1}(\tau)\,=\,0 \hspace{0.5cm} \text{if}\,\,\, i\in A \nonumber \\
& & arg f_{j2}(\tau)\,=\,0 \hspace{0.5cm} \text{if}\,\,\, j\in B.
\end{eqnarray}
Then one can use the components of the unit vector $\vec n_i$ to
parametrize the solution of the constraints Eq.(9)
\FL
\begin{eqnarray}
i\in A: \quad f_{i1} & = & \bar f_{i1}\,=\,\frac {1}{\sqrt 2}\,
\left(1\,+\,n_{i3}\right)^{\frac 12} \nonumber \\
f_{i2} & = & \frac {1}{\sqrt 2}
\frac {n^+_i}{\left(1\,+\,n_{i3}\right)^{\frac 12}}, \qquad\,\,
\bar f_{i2}\,=\,\frac {1}{\sqrt 2}
\frac {n^-_i}{\left(1\,+\,n_{i3}\right)^{\frac 12}} \nonumber \\
j\in B :\quad f_{j1} & = & \frac {1}{\sqrt 2}
\frac {n^-_j}{\left(1\,-\,n_{j3}\right)^{\frac 12}}, \qquad\,\,
\bar f_{j1}\,=\,\frac {1}{\sqrt 2}
\frac {n^+_j}{\left(1\,-\,n_{j3}\right)^{\frac 12}} \nonumber\\
f_{j2} & = & \bar f_{j2}\,=\,\frac {1}{\sqrt 2}\,
\left(1\,-\,n_{j3}\right)^{\frac 12}
\end{eqnarray}
where $n^{\pm}_r\,=\,n_{r1}\,\pm\,i\,n_{r2}$.

Now I am going to the derivation of the long-wavelength limit of the model.
To that purpose one introduces a new unit vector $\vec m_i\,
\left(\vec m_i^2\,=\,1\right)$ and a vector $\vec L_i$
\FL
\begin{eqnarray}
n_i & = & \,\,\,\sqrt {1\,-\,a^2 \vec L^2_i}\,\vec m_i\,+\,a\,\vec L_i, \qquad
\text{if}\quad i\in A \nonumber \\
n_j & = & -\,\sqrt {1\,-\,a^2\vec L^2_j}\,\vec m_j\,+\,a\,\vec L_j, \qquad
\text{if}\quad j\in B
\end{eqnarray}

The new spin-vector $\vec m_i$ is a smooth field on the lattice and $a$ is
the lattice spacing. The constraint $\vec n^2_i\,=\,1$ and the requirement
that the new vector $\vec m_i$ should obey the same constraint $\vec m_i^2\,=\,1$
demand $\vec m_i$ and $\vec L_i$ to be orthogonal
\FL
\begin{equation}
\vec m_i\cdot\vec L_i\,=\,0
\end{equation}

The next step is to substitute Eq.(15) into the fields $f_{i\sigma}(\tau)\,
\left(\bar f_{i\sigma}(\tau)\right)$ Eq.(14), and then to insert them into
the action. This yields an action, which depends on the vectors $\vec m_i,
\vec L_i$ and the fermionic fields. I expand the action in powers of the
vector $\vec L_i$, keeping only the first three terms in the expansion.

To begin with I adrress the terms with time derivatives
\FL
\begin{eqnarray}
S_{kin} & = & \int\limits^{\beta}_0 d\tau\sum\limits_i\left[\bar f_{i\sigma}(\tau)
\dot f_{i\sigma}(\tau)\left(1\,-\,\bar\psi_i(\tau)\psi_i(\tau)\right)\,+\,
\bar\psi_i(\tau)\dot\psi_i(\tau)\right] \nonumber \\
& = & \int\limits^{\beta}_0 d\tau\left\{{\sum\limits_{i\in A}} '
\left[\frac i2\,\vec {\cal A}(\vec n_i)\cdot\dot{\vec n_i}(\tau)
\left(1\,-\,\bar\psi^A_i(\tau)\psi^A_i(\tau)\right)\,+\,
\bar\psi^A_i(\tau)\dot\psi^A_i(\tau)\right]\right. \nonumber \\
& &\left.+\,{\sum\limits_{j\in B}} '
\left[\frac i2\,\vec {\cal A}(-\vec n_j)\cdot\dot{\vec n_j}(\tau)
\left(1\,-\,\bar\psi^B_j(\tau)\psi^B_j(\tau)\right)\,+\,
\bar\psi^B_j(\tau)\dot\psi^B_j(\tau)\right]\right\},
\end{eqnarray}
where
\FL
\begin{equation}
\frac i2\,\vec {\cal A}(\vec n_r)\,=\,\bar f_{r\sigma}(\vec n_r)\frac {\partial}
{\partial\vec n_r}\,f_{r\sigma}(\vec n_r)
\end{equation}
is the vector potential of a Dirac magnetic monopol at the center of the unit
sphere. It obeys locally
\FL
\begin{equation}
\vec\partial_{\vec n}\times\vec{\cal A}\left(\vec n \right)\,=\,\vec n
\end{equation}

Substituting (15) into (17) and keeping the terms up to order $a$ one obtains
\FL
\begin{eqnarray}
S_{kin} & = & \int\limits^{\beta}_0 d\tau \left\{\frac i2\,\sum\limits_i
(-1)^{|i|}\,\vec{\cal A}(\vec m)\cdot\dot{\vec m_i}(\tau) \right. \nonumber \\
& &\left. +\,\frac i2a\sum\limits_i\left(\frac {\partial\cal A^{\alpha}}
{\partial m_{i\beta}}\,L_{i\beta}(\tau)\dot m_{i\alpha}(\tau)\,+\,
\vec{\cal A}(\vec m)\cdot\dot{\vec L_i}(\tau)\right)\right. \nonumber \\
& & \left. +\,{\sum\limits_{i\in A}} '\bar\psi^A_i(\tau)\left[\partial_{\tau}\,
-\,\bar z_{i\sigma}(\tau)\dot z_{i\sigma}(\tau)\right]\psi^A_i(\tau)\,+\,
{\sum\limits_{j\in B}} '\bar\psi^B_j(\tau)\left[\partial_{\tau}\,+\,
\bar z_{j\sigma}(\tau)\dot z_{j\sigma}(\tau)\right]\psi^B_j(\tau)
\right. \nonumber \\
& & \left. -\frac i2a{\sum\limits_{i\in A}} '\bar\psi^A_i(\tau)\left[\frac
{\partial\cal A^{\alpha}(\vec m)}{\partial m_{i\beta}}\,L_{i\beta}(\tau)
\dot m_{i\alpha}(\tau)\,+\,\vec{\cal A}(\vec m)\cdot\dot{\vec L_i}(\tau)
\right]\psi^A_i(\tau)\right. \nonumber \\
& & \left. -\frac i2a{\sum\limits_{j\in B}} '\bar\psi^B_j(\tau)\left[\frac
{\partial\cal A^{\alpha}(\vec m)}{\partial m_{j\beta}}\,L_{j\beta}(\tau)
\dot m_{j\alpha}(\tau)\,+\,\vec{\cal A}(\vec m)\cdot\dot{\vec L_j}(\tau)
\right]\psi^B_j(\tau)\right\},
\end{eqnarray}
where I have introduced two complex fields $z_{i\sigma}(\tau)
\left(\bar z_{i\sigma}(\tau)\right)$
\FL
\begin{eqnarray}
z_{r1} & = & \bar z_{r1}\,=\,\frac {1}{\sqrt 2}\left(1\,+\,m_{r3}\right)^
{\frac 12} \nonumber \\
z_{r2} & = & \frac {1}{\sqrt 2}\frac {m^+_r}{\left(1\,+\,m_{r3}\right)^
{\frac 12}}\,\,,\qquad
\bar z_{r2}\,=\,\frac {1}{\sqrt 2}\frac {m^-_r}{\left(1\,+\,m_{r3}\right)
^{\frac 12}}\,\,,
\end{eqnarray}
which satisfy $\bar z_{r\sigma}z_{r\sigma}\,=\,1$ and $\vec m_r\,=\,\bar z_r
\vec \sigma z_r$.

The first term in Eq.(20) is not important in the two dimensional case and I
ignore it. The second term, after integration by parts, can be written in
the form
\FL
\begin{eqnarray}
& & \frac i2a\int\limits^{\beta}_0\sum\limits_i
\left(\frac {\partial\cal A^{\alpha}}
{\partial m_{i\beta}}\,L_{i\beta}(\tau)\dot m_{i\alpha}(\tau)\,+\,
\vec{\cal A}(\vec m)\cdot\dot{\vec L_i}(\tau)\right) \nonumber \\
& = & \frac i2a\int\limits^{\beta}_0\sum\limits_i
\left(\frac {\partial\cal A^{\alpha}}
{\partial m_{i\beta}}\,-\,\frac {\partial\cal A^{\beta}}
{\partial m_{i\alpha}}\right)\,L_{i\beta}(\tau)\dot m_{i\alpha}(\tau)
\nonumber \\
& = & \frac i2a\int\limits^{\beta}_0\sum\limits_i\epsilon_{\beta\gamma\alpha}
\left(\vec \partial_{\vec m}\times\vec{\cal A}\right)_{\gamma}\,
L_{i\beta}(\tau)\dot m_{i\alpha}(\tau) \nonumber \\
& = & \frac i2a\int\limits^{\beta}_0\sum\limits_i
\left(\vec m_i\times\dot{\vec m_i}\right)\cdot \vec L_i\,,
\end{eqnarray}
where the Eq.(19) is used.

The last two terms in Eq.(20) can be canceled by the transformation
\FL
\begin{equation}
\psi^R-r(\tau)\,\to e^{i\frac {a}{2}\Delta_r(\tau)}\,\psi^R_r(\tau)
\end{equation}
where
\FL
\begin{equation}
\dot\Delta_r(\tau)\,=\,\frac{\partial\cal A^{\alpha}(\vec m)}
{\partial m_{r\beta}}\,L_{r\beta}(\tau)\dot m_{r\alpha}(\tau)\,+\,
\vec{\cal A}(\vec m_r)\cdot\dot{\vec L_r}(\tau)
\end{equation}
and $R$ stands for $A$ or $B$. After this transformation phases appear
only in the hopping terms in the form of $\exp\left\{\frac {i}{2}a\left(
\Delta_r\,-\,\Delta_{r '}\right)\right\}$. In the continuum limit
$\Delta_r\,-\,\Delta_{r '}$ is of the order of $a$. Hence the phases give
no contribution to the effective action. This means, that in the long-
wavelength, low-frequency limit one can ignore the last two terms in Eq.(20).

Dealing with the hopping terms it is convenient to represent the vector
$\vec L_i$ in the form
\FL
\begin{equation}
\vec L_i\,=\,\bar\kappa_i\vec{\text{\large e}}_i\,+\,\kappa_i\vec{\bar{\text{
\large e}}}_i
\end{equation}
where the complex vectors $\vec{\text{\large e}}_i$ and the conjugated vector
$\vec{\bar{\text{\large e}}}_i$ are orthogonal to the vector $\vec m_i$ and
satisfy
\FL
\begin{equation}
\vec {\text{\large e}}^2_i\,=\,\vec {\bar {\text {\large e}}}^2_i\,=\,0\,\,,
\hspace{1.5cm}
\vec {\bar {\text {\large e}}}_i\cdot\vec {\text {\large e}}_i\,=\,\frac 12.
\end{equation}
The explicit expression for the vectors are
\FL
\begin{eqnarray}
& & \text{\large e}_{i1}\,=\,\frac 12\,\left(z_{i1}z_{i1}\,-\,z_{i2}z_{i2}
\right) \hspace{1cm}
\bar{\text{\large e}}_{i1}\,=\,\frac 12\,\left(\bar z_{i1}\bar z_{i1}\,-\,
\bar z_{i2}\bar z_{i2}\right) \nonumber \\
& & \text{\large e}_{i2}\,=\,\frac i2\,\left(z_{i1}z_{i1}\,+\,z_{i2}z_{i2}
\right)
\hspace{1cm}
\bar{\text{\large e}}_{i2}\,=\,\frac {1}{2i}\,
\left(\bar z_{i1}\bar z_{i1}\,+\,\bar z_{i2}\bar z_{i2}\right) \nonumber \\
& & \text{\large e}_{i3}\,=\,-\,z_{i1}z_{i2}
\hspace{3cm}
\bar{\text{\large e}}_{i3}\,=\,-\,\bar z_{i1}\bar z_{i2}
\end{eqnarray}

The fields $f_{i\sigma}(\tau)\,\left(\bar f_{i\sigma}(\tau)\right)$ depend
on the vectors $\vec m_i(\tau)$ and the fields $\kappa_i(\tau)\,
\left(\bar\kappa_i(\tau)\right)$. I expand them in powers of $\kappa_i(\tau)$
and $\bar\kappa_i(\tau)$ up to linear terms. This yields
\FL
\begin{eqnarray}
& & f_{i\sigma}\bar f_{j\sigma}\,\simeq\,-\,z_{i1}z_{j2}\,+\,z_{i2}z_{j1}\,+\,
a\,\kappa_i \nonumber \\
& & f_{j\sigma}\bar f_{i\sigma}\,\simeq\,-\,\bar z_{i1}\bar z_{j2}\,+\,
\bar z_{i2}\bar z_{j1}\,+\,a\,\bar\kappa_i ,
\end{eqnarray}
if $i\in A$ and $j\,=\,i\,+\,a_{\mu}$
\FL
\begin{eqnarray}
& & f_{j\sigma}\bar f_{i\sigma}\,\simeq\,-\,\bar z_{j2}\bar z_{i1}\,+\,
\bar z_{j1}\bar z_{i2}\,+\,a\,\bar\kappa_j ,\nonumber \\
& & f_{i\sigma}\bar f_{j\sigma}\,\simeq\,-\,z_{j2}z_{i1}\,+\,z_{j1}z_{i2}\,+\,
a\,\kappa_j
\end{eqnarray}
if $j\in B$ and $i\,=\,j\,+\,a_{\mu}$
\FL
\begin{eqnarray}
& & f_{i\sigma}\bar f_{j\sigma}\,\simeq\,1\,+\,z_{i\sigma}\left(\bar z_{j\sigma}
\,-\,\bar z_{i\sigma}\right) \nonumber \\
& & f_{j\sigma}\bar f_{i\sigma}\,\simeq\,1\,+\,\bar z_{i\sigma}
\left(z_{j\sigma}\,-\,z_{i\sigma}\right)
\end{eqnarray}
if $i,j\in A$ and $j\,=\,i\,+\,e_{\lambda}$
\FL
\begin{eqnarray}
& & f_{j\sigma}\bar f_{i\sigma}\,\simeq\,1\,+\,\bar z_{j\sigma}
\left(z_{i\sigma}\,-\,z_{j\sigma}\right) \nonumber \\
& & f_{i\sigma}\bar f_{j\sigma}\,\simeq\,1\,+\,z_{j\sigma}
\left(\bar z_{i\sigma}\,-\,\bar z_{j\sigma}\right)
\end{eqnarray}
if $j,i\in B$ and $i\,=\,j\,+\,e_{\lambda}$. The two lattice's directions
$(a,0)$ and $(0,a)$ are noted by $a_{\mu},\mu=x,y$, and $e_{\lambda}\,=\,
\left[a_x\,+\,a_y\,,\,a_x\,-\,a_y\right]$. I have used again the two complex
fields defined by Eq.(21).

Collecting the results above, one can write the action in the form
\FL
\begin{equation}
S\,=\,S_0\,+\,S_L\,+\,S_{LL}
\end{equation}
The term which does not depend on $\vec L$ reads
\FL
\begin{eqnarray}
S_0\,=\,& & \int\limits^{\beta}_0 d\tau\left\{{\sum\limits_{i\in A}} '
\bar\psi^A_i(\tau)\left[\partial_{\tau}\,-\,\bar z_{i\sigma}\dot z_{i\sigma}
\right]\psi^A_i(\tau)\,+\,{\sum\limits_{j\in B}} '
\bar\psi^B_j(\tau)\left[\partial_{\tau}\,+\,\bar z_{j\sigma}\dot z_{j\sigma}
\right]\psi^B_j(\tau)\right. \nonumber \\
& & \left.+\,\frac J2\,{\sum\limits_{i\in A,\mu}} '
\left[\bar\psi^A_i(\tau)\psi^A_i(\tau)\,+\,\bar\psi^B_{i+a_{\mu}}(\tau)
\psi^B_{i+a_{\mu}}(\tau)\right]\right. \nonumber \\
& & \left.+\,\frac J2\,{\sum\limits_{j\in B,\mu}} '
\left[\bar\psi^B_j(\tau)\psi^B_j(\tau)\,+\,\bar\psi^A_{j+a_{\mu}}(\tau)
\psi^A_{j+a_{\mu}}(\tau)\right]\right. \nonumber \\
& & \left. -\,t_2\,{\sum\limits_{i\in A,\lambda}} '
\left[\bar\psi^A_i(\tau)\psi^A_{i+e_{\lambda}}(\tau)\,+\,h.c.\right]\,
-\,t_2\,{\sum\limits_{j\in B,\lambda}} '
\left[\bar\psi^B_j(\tau)\psi^B_{j+e_{\lambda}}(\tau)\,+\,h.c.\right]\right.
\nonumber \\
& & \left.-\,t_1\,{\sum\limits_{i\in A,\mu}} '
\left[\bar\psi^A_i(\tau)\psi^B_{i+a_{\mu}}(\tau)\left(-z_{i1}(\tau)
z_{i+a_{\mu}\,2}(\tau)\,+\,z_{i2}(\tau)z_{i+a_{\mu}\,1}(\tau)\right)\,
+\,h.c.\right]\right. \nonumber \\
& & \left.-\,t_1\,{\sum\limits_{j\in B,\mu}} '
\left[\bar\psi^B_j(\tau)\psi^A_{i+a_{\mu}}(\tau)\left(-\bar z_{j2}(\tau)
\bar z_{j+a_{\mu}\,1}(\tau)\,+\,\bar z_{j1}(\tau)\bar z_{j+a_{\mu}\,2}(\tau)
\right)\,+\,h.c.\right]\right. \nonumber \\
& & \left.-\,t_2\,{\sum\limits_{i\in A,\lambda}} '
\left[\bar\psi^A_i(\tau)\psi^A_{i+e_{\lambda}}(\tau)z_{i\sigma}(\tau)
\left(\bar z_{i+e_{\lambda}\,\sigma}(\tau)\,-\,\bar z_{i\sigma}(\tau)
\right)\,+\,h.c.\right]\right. \nonumber \\
& & \left.-\,t_2\,{\sum\limits_{j\in B,\lambda}} '
\left[\bar\psi^B_j(\tau)\psi^B_{j+e_{\lambda}}(\tau)\bar z_{j\sigma}(\tau)
\left(z_{j+e_{\lambda}\,\sigma}(\tau)\,-\,z_{j\sigma}(\tau)
\right)\,+\,h.c.\right]\right. \nonumber \\
& & \left.+\,\frac J8\,{\sum\limits_{i\in A,\mu}} '
\left(\vec m_{i+a_{\mu}}(\tau)\,-\,\vec m_i(\tau)\right)^2
\left(1\,-\,\bar\psi^A_i(\tau)\psi^A_i(\tau)\right)
\left(1\,-\,\bar\psi^B_{i+a_{\mu}}(\tau)\psi^B_{i+a_{\mu}}(\tau)\right)
\right. \nonumber \\
& & \left.+\,\frac J8\,{\sum\limits_{j\in B,\mu}} '
\left(\vec m_{j+a_{\mu}}(\tau)\,-\,\vec m_j(\tau)\right)^2
\left(1\,-\,\bar\psi^B_j(\tau)\psi^B_j(\tau)\right)
\left(1\,-\,\bar\psi^A_{j+a_{\mu}}(\tau)\psi^A_{j+a_{\mu}}(\tau)\right)
\right. \nonumber \\
& & \left.-\,\frac J2\,{\sum\limits_{i\in A,\mu}} '
\bar\psi^A_i(\tau)\psi^A_i(\tau)
\bar\psi^B_{i+a_{\mu}}(\tau)\psi^B_{i+a_{\mu}}(\tau)
-\,\frac J2\,{\sum\limits_{j\in B,\mu}} '
\bar\psi^B_j(\tau)\psi^B_j(\tau)
\bar\psi^A_{j+a_{\mu}}(\tau)\psi^A_{j+a_{\mu}}(\tau)\right. \nonumber \\
& & \left.-\,\mu\,{\sum\limits_{i\in A}} '\left(1\,-\,\bar\psi^A_i(\tau)
\psi^A_i(\tau)\right)
-\,\mu\,{\sum\limits_{j\in B}} '\left(1\,-\,\bar\psi^B_j(\tau)
\psi^B_j(\tau)\right)\right\}
\end{eqnarray}

It is convenient to write the linear term in the form
\FL
\begin{equation}
S_L\,=\,a\,\int\limits^{\beta}_0 d\tau\left[{\sum\limits_{i\in A}} '\left(
\bar\kappa_i\rho^A_i\,+\,\kappa_i\tilde\rho^A_i\right)\,+\,
{\sum\limits_{j\in B}} '\left(
\bar\kappa_j\rho^B_j\,+\,\kappa_j\tilde\rho^B_j\right)\right]
\end{equation}
where
\FL
\begin{eqnarray}
\rho^A_i & = & \frac i2 \left(\vec m_i\times\dot{\vec m_i}\right)\cdot
\vec{\text{\large e}}_i\,-\,t_1\sum\limits_{\mu}\bar\psi^B_{i+a_{\mu}}\psi^A_i
\nonumber \\
\tilde\rho^A_i & = & \frac i2 \left(\vec m_i\times\dot{\vec m_i}\right)\cdot
\vec{\bar{\text{\large e}}}_i\,-\,t_1\sum\limits_{\mu}\bar\psi^A_i\psi^B_{i+
a_{\mu}}
\nonumber \\
\rho^B_j & = & \frac i2 \left(\vec m_j\times\dot{\vec m_j}\right)\cdot
\vec{\text{\large e}}_j\,-\,t_1\sum\limits_{\mu}\bar\psi^B_j\psi^A_{j+a_{\mu}}
\nonumber \\
\tilde\rho^B_j & = & \frac i2 \left(\vec m_j\times\dot{\vec m_j}\right)\cdot
\vec{\bar{\text{\large e}}}_j\,-\,t_1\sum\limits_{\mu}\bar\psi^A_{j+a_{\mu}}
\psi^B_j
\end{eqnarray}
Finally, the bilinear term is
\FL
\begin{eqnarray}
S_{LL}\,=\,\frac {Ja^2}{2}\,\int\limits^{\beta}_0\, d
& \tau &
\left[{\sum\limits_{i\in A}} '\vec L^2_i
\left(1\,-\,\bar\psi^A_i\psi^A_i\right)
\left(1\,-\,\bar\psi^B_{i+a_{\mu}}\psi^B_{i+a_{\mu}}\right)\right. \nonumber \\
& & \left. +\,{\sum\limits_{j\in B}} '\vec L^2_j\left(1\,-\,\bar\psi^B_j\psi^B_j\right)
\left(1\,-\,\bar\psi^A_{j+a_{\mu}}\psi^A_{j+a_{\mu}}\right)\right]
\end{eqnarray}
where $\vec L^2_i\,=\,\bar\kappa_i\kappa_i$.

The last step is to integrate over the vector $\vec L_i$. The integral over
$\vec L_i$ is defined as an integral over the independent variables
$\bar\kappa_i(\tau)$ and $\kappa_i(\tau)$. Carring out the integration, one
obtains the action of the effective theory.
\FL
\begin{equation}
S_{eff}\,=\,S_{CP^1}\,+\,S_F
\end{equation}
where $S_{CP^1}$ is the action of the $CP^1$ model ($\sigma$-model) and
$S_F$ is the part of the effective action which depends on the vector $\vec m_i$
(complex fields $z_{i\sigma},\bar z_{i\sigma}$) and the fermionic fields
\FL
\begin{eqnarray}
S_F\,=\,& & \int\limits^{\beta}_0 d\tau\left\{{\sum\limits_{i\in A}} '
\bar\psi^A_i(\tau)\left[\partial_{\tau}\,-\,\bar z_{i\sigma}\dot z_{i\sigma}
\right]\psi^A_i(\tau)\,+\,{\sum\limits_{j\in B}} '
\bar\psi^B_j(\tau)\left[\partial_{\tau}\,+\,\bar z_{j\sigma}\dot z_{j\sigma}
\right]\psi^B_j(\tau)\right. \nonumber \\
& & \left.+\,\frac J2\,{\sum\limits_{i\in A,\mu}} '
\left[\bar\psi^A_i(\tau)\psi^A_i(\tau)\,+\,\bar\psi^B_{i+a_{\mu}}(\tau)
\psi^B_{i+a_{\mu}}(\tau)\right]\right. \nonumber \\
& & \left.+\,\frac J2\,{\sum\limits_{j\in B,\mu}} '
\left[\bar\psi^B_j(\tau)\psi^B_j(\tau)\,+\,\bar\psi^A_{j+a_{\mu}}(\tau)
\psi^A_{j+a_{\mu}}(\tau)\right]\right. \nonumber \\
& & \left. -\,t_2\,{\sum\limits_{i\in A,\lambda}} '
\left[\bar\psi^A_i(\tau)\psi^A_{i+e_{\lambda}}(\tau)\,+\,h.c.\right]\,
-\,t_2\,{\sum\limits_{j\in B,\lambda}} '
\left[\bar\psi^B_j(\tau)\psi^B_{j+e_{\lambda}}(\tau)\,+\,h.c.\right]\right.
\nonumber \\
& & \left. +\,i\frac {t_1}{J}{\sum\limits_{i\in A,\mu}} '
\left(\vec m_i\times\dot{\vec m}_i\right)\cdot\left[
\vec{\bar{\text{\large e}}}_{i,i+a_{\mu}}\bar\psi^B_{i+a_{\mu}}\psi^A_i\,+
\vec{\text{\large e}}_{i,i+a_{\mu}}\bar\psi^A_i\psi^B_{i+a_{\mu}}\right]
\right. \nonumber \\
& & \left. +\,i\frac {t_1}{J}{\sum\limits_{j\in B,\mu}} '
\left(\vec m_j\times\dot{\vec m}_j\right)\cdot\left[
\vec{\bar{\text{\large e}}}_{j,j+a_{\mu}}\bar\psi^B_j\psi^A_{j+a_{\mu}}\,+
\vec{\text{\large e}}_{j,j+a_{\mu}}\bar\psi^A_{j+a_{\mu}}\psi^B_j\right]
\right. \nonumber \\
& & \left.-\,t_1\,{\sum\limits_{i\in A,\mu}} '
\left[\bar\psi^A_i(\tau)\psi^B_{i+a_{\mu}}(\tau)\left(-z_{i1}(\tau)
z_{i+a_{\mu}\,2}(\tau)\,+\,z_{i2}(\tau)z_{i+a_{\mu}\,1}(\tau)\right)\,
+\,h.c.\right]\right. \nonumber \\
& & \left.-\,t_1\,{\sum\limits_{j\in B,\mu}} '
\left[\bar\psi^B_j(\tau)\psi^A_{i+a_{\mu}}(\tau)\left(-\bar z_{j2}(\tau)
\bar z_{j+a_{\mu}\,1}(\tau)\,+\,\bar z_{j1}(\tau)\bar z_{j+a_{\mu}\,2}(\tau)
\right)\,+\,h.c.\right]\right. \nonumber \\
& & \left.-\,t_2\,{\sum\limits_{i\in A,\lambda}} '
\left[\bar\psi^A_i(\tau)\psi^A_{i+e_{\lambda}}(\tau)z_{i\sigma}(\tau)
\left(\bar z_{i+e_{\lambda}\,\sigma}(\tau)\,-\,\bar z_{i\sigma}(\tau)
\right)\,+\,h.c.\right]\right. \nonumber \\
& & \left.-\,t_2\,{\sum\limits_{j\in B,\lambda}} '
\left[\bar\psi^B_j(\tau)\psi^B_{j+e_{\lambda}}(\tau)\bar z_{j\sigma}(\tau)
\left(z_{j+e_{\lambda}\,\sigma}(\tau)\,-\,z_{j\sigma}(\tau)
\right)\,+\,h.c.\right]\right. \nonumber \\
& & \left.+{\sum\limits_{i\in A,\mu}} '
\left[\frac {1}{32J}\dot{\vec m}_i\cdot\dot{\vec m}_i-
\frac {J}{8}\left(\vec m_{i+a_{\mu}}(\tau)-\vec m_i(\tau)\right)^2\right]
\left(\bar\psi^A_i(\tau)\psi^A_i(\tau)
+\bar\psi^B_{i+a_{\mu}}(\tau)\psi^B_{i+a_{\mu}}(\tau)\right)
\right. \nonumber \\
& & \left.+{\sum\limits_{j\in B,\mu}} '
\left[\frac {1}{32J}\dot{\vec m}_j\cdot\dot{\vec m}_j-
\frac {J}{8}\left(\vec m_{j+a_{\mu}}(\tau)-\vec m_j(\tau)\right)^2\right]
\left(\bar\psi^B_j(\tau)\psi^B_j(\tau)
+\bar\psi^A_{j+a_{\mu}}(\tau)\psi^A_{j+a_{\mu}}(\tau)\right)
\right. \nonumber \\
& & \left.+\,\frac {\lambda}{4}\,{\sum\limits_{i\in A,\mu}} '
\bar\psi^A_i(\tau)\psi^A_i(\tau)
\bar\psi^B_{i+a_{\mu}}(\tau)\psi^B_{i+a_{\mu}}(\tau)
+\,\frac {\lambda}{4}\,{\sum\limits_{j\in B,\mu}} '
\bar\psi^B_j(\tau)\psi^B_j(\tau)
\bar\psi^A_{j+a_{\mu}}(\tau)\psi^A_{j+a_{\mu}}(\tau)\right. \nonumber \\
& & \left.-\,\mu\,{\sum\limits_{i\in A}} '\left(1\,-\,\bar\psi^A_i(\tau)
\psi^A_i(\tau)\right)
-\,\mu\,{\sum\limits_{j\in B}} '\left(1\,-\,\bar\psi^B_j(\tau)
\psi^B_j(\tau)\right)\right\}\,+\,S_{add},
\end{eqnarray}
where
\FL
\begin{equation}
\frac {\lambda}{4}\,=\,\frac {t^2_1}{J}\,-\,\frac J2
\end{equation}
I have replaced in Eq.(38) $\vec{\text{\large e}}_r$ with
$\vec{\text{\large e}}_{r,r'}$ and $\vec{\bar{\text{\large e}}}_r$ with
$\vec{\bar{\text{\large e}}}_{r,r'}$ where
\FL
\begin{eqnarray}
& & \text{\large e}_{rr'1}\,=\,\frac 12\,\left(z_{r1}z_{r'1}\,-\,z_{r2}z_{r'2}
\right) \hspace{1cm}
\bar{\text{\large e}}_{rr'1}\,=\,\frac 12\,\left(\bar z_{r1}\bar z_{r'1}\,-\,
\bar z_{r2}\bar z_{r'2}\right) \nonumber \\
& & \text{\large e}_{rr'2}\,=\,\frac i2\,\left(z_{r1}z_{r'1}\,+\,z_{r2}z_{r'2}
\right)
\hspace{1cm}
\bar{\text{\large e}}_{rr'2}\,=\,\frac {1}{2i}\,
\left(\bar z_{r1}\bar z_{r'1}\,+\,\bar z_{r2}\bar z_{r'2}\right) \nonumber \\
& & \text{\large e}_{rr'3}\,=\,-\,\frac 12\left(z_{r1}z_{r'2}\,+\,
z_{r2}z_{r'1}\right)
\hspace{0.8cm}
\bar{\text{\large e}}_{rr'3}\,=\,-\,\frac 12\left(\bar z_{r1}\bar z_{r'2}\,+\,
\bar z_{r2}\bar z_{r'1}\right)
\end{eqnarray}
The difference is of order $a$ and it does not affect the long-wavelength
physics.

The additional action $S_{add}$ contains all terms in higher order of
derivatives and fields. They do not contribute to the long-wavelength
physics, and hereafter I shall not consider them.

Until now the fields $z_{i\sigma}\left(\bar z_{i\sigma}\right)$ have been
viewed as defined by Eqs.(21). Now, I consider
$z_{i\sigma}\left(\bar z_{i\sigma}\right)$ as independent bose fields which
satisfy the constraint $\bar z_{i\sigma}z_{i\sigma}\,=\,1$ and the spin vector
$\vec m_i$ as defined by the equality $\vec m_i\,=\,\bar z_i\vec\sigma z_i$.
Then, the action (38) is invariant under the $U(1)$ gauge transformations
\FL
\begin{eqnarray}
& & z'_{r\sigma}(\tau)\,=\,e^{i\alpha_r(\tau)}z_{r\sigma}(\tau) ;\,\,
\,\,\hspace{1cm}
\bar z'_{r\sigma}(\tau)\,=\,e^{-i\alpha_r(\tau)}\bar z_{r\sigma}(\tau)
\nonumber \\
& & \psi '^A_r(\tau)\,=\,e^{i\alpha_r(\tau)}\psi^A_r(\tau)\,\,\,\text{if}\,\,\,
r\in A \nonumber \\
& & \psi '^B_r(\tau)\,=\,e^{-i\alpha_r(\tau)}\psi^B_r(\tau)\,\,\,\text{if}\,\,\,
r\in B.
\end{eqnarray}
One can restore the representation (21) of the fields imposing the gauge-fixing
condition $arg z_{r1}\,=\,0$.

An important point in the effective model (38) is the four-fermion term. In
the starting Hamiltonian (11) the four-fermion interaction is attractive. This,
sometimes, leads to a speculative conjecture about superconductivity. But the
sign in front of the four-fermion term in Eq.(11) is just an output of the
parametrization. An additional repulsive four-fermion interaction appears in the
effective theory (38) due to the interaction of the fermions with the "fast
modes" of the spinon $\left(\vec L_i\right)$. For the parameter range
$\lambda > 0$, it screens the attractive four-fermion interaction. I shall
return to this term in the next section.

The effective theory Eq.(38) is a theory of slow spinon modes defined on a small
area around the zero vector, and fermions defined on a whole antiferromagnetic
Brillouin zone. All fermionic terms are taken into account exept for those of
order equal or higher then six. This permits to investigate more special phases,
related to the geometry of the lattice.

To carry out the long-wavelength limit for fermions, one should know the exact
Fermi surface. But for small doping, it is enough to consider the dispersion
of free fermions. In the model Eq.(38) with $t_2\,>\,0$, it has minima located
at zero wave-vector, and the continuum limit can be achieved by means of a
gradient-expansion around this point. In this way one obtains the following
continuum theory
\FL
\begin{eqnarray}
S_{eff}\,= & & \int d^2x d \tau \left\{\frac {2}{g^2}\left[
\overline{D_{\tau}z}_{\sigma}D_{\tau}z_{\sigma}\,+\,c^2\,
\overline{D_{\mu}z}_{\sigma}D_{\mu}z_{\sigma}\right]\right. \nonumber \\
& & \left.+\,\bar\psi^A D^{(A)}_{\tau}\psi^A\,+\,\frac {1}{2m}
\overline{D^{(A)}_{\mu}\psi^A}D^{(A)}_{\mu}\psi^A
\,+\,\bar\psi^B D^{(B)}_{\tau}\psi^B\,+\,\frac {1}{2m}
\overline{D^{(B)}_{\mu}\psi^B}D^{(B)}_{\mu}\psi^B \right. \nonumber \\
& & \left.-\,\frac {2t_1}{J}\,\bar\psi^A\psi^B\left(z_1\dot z_2\,-\,z_2\dot z_1
\right)\,+\,\frac {2t_1}{J}\,\bar\psi^B\psi^A\left(\bar z_1\dot{\bar z}_2\,-\,
\bar z_2\dot{\bar z}_1\right)\right. \nonumber \\
& & \left. +\,t_1\,a^2\left(\bar\psi^A\partial_{\mu}\psi^B\,-\,\partial_{\mu}
\bar\psi^A\psi^B\right)
\left(z_1\partial_{\mu}z_2\,-\,z_2\partial_{\mu}z_1\right) \right. \nonumber \\
& & \left.-\,t_1\,a^2\left(\bar\psi^B\partial_{\mu}\psi^A\,-\,\partial_{\mu}
\bar\psi^B\psi^A\right)
\left(\bar z_1\partial_{\mu}\bar z_2\,-\,\bar z_2\partial_{\mu}\bar z_1\right)
\right. \nonumber \\
& & \left.+\,\frac {2}{\tilde g}
\left[\overline{D_{\tau}z}_{\sigma}D_{\tau}z_{\sigma}
\,+\,\tilde c^2 \overline{D_{\mu}z}_{\sigma}D_{\mu}z_{\sigma}\right]
\left(\bar\psi^A\psi^A\,+\,\bar\psi^B\psi^B\right)\right. \nonumber \\
& & \left.+\,\frac {\lambda a^2}{2}\bar\psi^A\psi^A\bar\psi^B\psi^B\,+
\,\mu\,\left(\bar\psi^A\psi^A\,+\,\bar\psi^B\psi^B\right)
\right\},
\end{eqnarray}
where
\FL
\begin{eqnarray}
& & D_l\,z_{\sigma}\,=\,\left(\partial_l\,-\,\bar z_{\sigma '}\partial_l
z_{\sigma '}\right)z_{\sigma}, \,\,\,\,\,\,l\,=\,0,x,y \nonumber \\
& & D^{(A)}_l\,\psi^A\,=\,\left(\partial_l\,-\,\bar z_{\sigma '}\partial_l
z_{\sigma '}\right)\psi^A, \,\,\,\,\,\,
D^{(B)}_l\,\psi^B\,=\,\left(\partial_l\,+\,\bar z_{\sigma '}\partial_l
z_{\sigma '}\right)\psi^B,
\end{eqnarray}
and the parameters are given by the equalities:\,\,\,\,\,$g=2a\sqrt J,\,\,\,\,
\,\,c=aJ,\,\,\,\,\tilde g=2\sqrt J,\,\,\,\,\,\,
\tilde c^2=2a^2J\left(2t_2\,-\,J\right)$\,\,\,\, and\,\,\,\,
$\frac {1}{2m}=2t_2a^2$.

To obtain the effective action Eq.(42) I have rescaled the Fermi fields
$\frac 1a\psi^R \to \psi^R,\,\,(R=A\text{or}B)$ and have used the identities
\FL
\begin{eqnarray}
&\left(\vec m\times\partial_l\vec m\right)\,\cdot\vec{\bar{\text{\large e}}}
\,\,& =  -i\,\left(\bar z_1\partial_l\bar z_2\,-\,\bar z_2\partial_l
\bar z_1\right) \nonumber \\
&\left(\vec m\times\partial_l\vec m\right)\,\cdot\vec{\text{\large e}}
& = i\,\left(z_1\partial_l z_2\,-\,z_2\partial_l z_1\right) \nonumber \\
&\frac 14 \partial_l\vec m\cdot\partial_l\vec m & =
\partial_l\bar z_{\sigma}\partial_l z_{\sigma}\,+\,\frac 14
\left(\bar z_{\sigma}\partial_l z_{\sigma}\,-
\,z_{\sigma}\partial_l\bar z_{\sigma}\right)^2 \nonumber \\
& & =  \left(\bar z_1\partial_l\bar z_2\,-
\,\bar z_2\partial_l \bar z_1\right)\left(z_1\partial_l z_2\,-
\,z_2\partial_l z_1\right),
\end{eqnarray}
where $l$ stands for $\tau, x,$ or $y$ and no sum over $l$ is assumed.
\ \\
\ \\
\ \\

{\bf III Gauge symmetry breaking}

The four-fermion terms in the effective action allow an appropriate mean-field
theory of gauge symmetry breaking. To demonstrate this I arrange the Fermi
fields in the form
\FL
\begin{equation}
S_{F^4}\,=\,\int\limits^{\beta}_0 d\tau\left\{-\frac {\lambda}{4}
{\sum\limits_{i\in A\mu}} '\bar\psi^A_i\psi^B_{i+a_{\mu}}\bar\psi^B_{i+a_{\mu}}
\psi^A_i\,-\,\frac {\lambda}{4}
{\sum\limits_{j\in B\mu}} '\bar\psi^B_j\psi^A_{j+a_{\mu}}\bar\psi^A_{j+a_{\mu}}
\psi^B_j\right\}.
\end{equation}
Then, by means of the Hubbard-Stratanovich transformation I introduce new
collective complex fields $u^A_{i\mu}(\tau),\,\bar u^A_{i\mu}(\tau),\,
u^B_{j\mu}(\tau),\,\bar u^B_{j\mu}(\tau)$, and rewrite the exponent in the form
\FL
\begin{eqnarray}
& & e^{-S_{F^4}}\,=\,\int \prod\limits_{i\mu\tau}d\bar u^A_{i\mu}(\tau)
du^A_{i\mu}(\tau)\prod\limits_{j\mu\tau} d\bar u^B_{j\mu}(\tau)
du^B_{j\mu}(\tau)\,
exp\int\limits^{\beta}_0 d\tau\left\{\frac {\lambda}{4}
{\sum\limits_{i\in A\mu}} '
\left[\bar u^A_{i\mu}(\tau)u^A_{i\mu}(\tau)\right.\right. \nonumber \\
& & \left.\left.-\bar u^A_{i\mu}\bar
\psi^B_{i+a_{\mu}}(\tau)\psi^A_i(\tau)-\bar\psi^A_i(\tau)
\psi^B_{i+a_{\mu}}(\tau)u^A_{i\mu}(\tau)\right] \right. \nonumber \\
& & \left.+\frac {\lambda}{4} {\sum\limits_{j\in B\mu}} '
\left[\bar u^B_{j\mu}(\tau)u^B_{j\mu}(\tau)\,-\,\bar u^B_{j\mu}\bar
\psi^A_{j+a_{\mu}}(\tau)\psi^B_j(\tau)\,-\,\bar\psi^B_j(\tau)
\psi^A_{j+a_{\mu}}(\tau)u^B_{j\mu}(\tau)\right] \right\}
\end{eqnarray}

The mean-field approximation  for the problem is just the evaluation
of the path integral over the new collective fields by means of the
saddle-point approximation. The stationary conditions are
\FL
\begin{equation}
\frac {\delta\cal F}{\delta u^A_{i\mu}}\,=\,0,\,
\frac {\delta\cal F}{\delta \bar u^A_{i\mu}}\,=\,0,\,
\frac {\delta\cal F}{\delta u^B_{j\mu}}\,=\,0,\,
\frac {\delta\cal F}{\delta \bar u^B_{j\mu}}\,=\,0,
\end{equation}
where
\FL
\begin{equation}
\cal F\,=\,-\,\frac {\lambda}{4\beta N_1}\,\int\limits^{\beta}_0 d\tau
\left[{\sum\limits_{i\in A,\mu}}'\bar u^A_{i\mu}(\tau) u^A_{i\mu}(\tau)
\,+\,{\sum\limits_{j\in B,\mu}}'\bar u^B_{j\mu}(\tau) u^B_{j\mu}(\tau)\right]\,
+\,\cal F_0
\end{equation}
and $\cal F_0$ is the free energy of a system with Hamiltonian
\FL
\begin{eqnarray}
h_{m.f.}\,= & - &t_2{\sum\limits_{i\in A\lambda}}'\left[\bar\psi^A_i
\psi^A_{i+e_{\lambda}}\,+\,h.c\right]\,-\,t_2{\sum\limits_{j\in B\lambda}}'
\left[\bar\psi^B_j\psi^B_{j+e_{\lambda}}\,+\,h.c\right] \nonumber \\
& + & \frac {\lambda}{4}{\sum\limits_{i\in A\mu}}'
\left[\bar u^A_{i\mu}\bar \psi^B_{i+a_{\mu}}(\tau)\psi^A_i(\tau)\,+\,\bar\psi^A_i(\tau)
\psi^B_{i+a_{\mu}}(\tau)u^A_{i\mu}(\tau)\right] \nonumber \\
& + & \,\frac {\lambda}{4}{\sum\limits_{j\in B\mu}}'
\left[\bar u^B_{j\mu}\bar
\psi^A_{j+a_{\mu}}(\tau)\psi^B_j(\tau)\,+\,\bar\psi^B_j(\tau)
\psi^A_{j+a_{\mu}}(\tau)u^B_{j\mu}(\tau)\right] \nonumber \\
& + & \mu {\sum\limits_{i\in A\mu}}'\bar\psi^a_i\psi^A_i\,+\,
\mu {\sum\limits_{j\in B\mu}}'\bar\psi^B_j\psi^B_j
\end{eqnarray}

The mean field equations (47) have a trivial solution $u^A\,=\,\bar u^A\,=\,
u^B\,=\,\bar u^B\,=\,0$, but, when $\lambda >0$ they have and non-zero
solution which leads to the breaking of the gauge symmetry.

In the phase with broken gauge symmetry the normal Green functions read
\FL
\begin{equation}
S^{AA}_k(\tau-\tau ')\,=\,S^{BB}_k(\tau-\tau ')\,=\,\frac {1}{\beta}
\sum\limits_{\omega_n}\,e^{i\omega_n(\tau-\tau')}
\frac {i\omega_n+\varepsilon_k}
{(i\omega_n+\varepsilon_k)^2\,-\,|\gamma_k|^2}
\end{equation}
and for the anomalous Green functions one obtains
\FL
\begin{eqnarray}
S^{BA}_k(\tau-\tau ') & = & \frac {1}{\beta}
\sum\limits_{\omega_n}\,e^{i\omega_n(\tau-\tau')}
\frac {\gamma_k}{(i\omega_n+\varepsilon_k)^2\,-\,|\gamma_k|^2}
\nonumber \\
S^{AB}_k(\tau-\tau ') & = & \frac {1}{\beta}
\sum\limits_{\omega_n}\,e^{i\omega_n(\tau-\tau')}
\frac {\bar\gamma_k}{(i\omega_n+\varepsilon_k)^2\,-\,|\gamma_k|^2}
\end{eqnarray}
The sum is over the frequencies $\omega_n=(2n+1)\frac{\pi}{\beta},\,
\varepsilon=\mu-4t_2\cos k_x\cos k_y$,\,and $\gamma_k=\bar u^A_{\mu}
e^{-iak_{\mu}}\,+\,u^B_{\mu}e^{iak_{\mu}}$ where the order parameters
$u^A_{\mu}$ and $u^B_{\mu}$ are choosen to be homogeneous.

Integrating over the fermions around the mean-field values of the order
parameters one obtains an effective action which is not gauge invariant
\FL
\begin{eqnarray}
S_{eff}'\,= & & \int d^2x d \tau \left\{\frac {2}{g^2_r}\left[
\overline{D_{\tau}z}_{\sigma}D_{\tau}z_{\sigma}\,+\,c^2_r\,
\overline{D_{\mu}z}_{\sigma}D_{\mu}z_{\sigma}\right]\right. \nonumber \\
& &\left.+\,\bar W_l\left(z_1\partial_lz_2\,-\,z_2\partial_lz_1
\right)\,+\,
W_l\left(\bar z_1\partial_l\bar z_2\,-\,\bar z_2\partial_l\bar z_1
\right) \right. \nonumber \\
& & \left.+\, Z_l\left[\left(z_1\partial_{\mu}z_2\,-\,z_2\partial_{\mu}z_1
\right)^2\,+\,\left(\bar z_1\partial_{\mu}\bar z_2\,-
\,\bar z_2\partial_{\mu}\bar z_1\right)^2\right] \right. \nonumber \\
& & \left.+\, M_l\left(\bar z_{\sigma}\partial_l z_{\sigma}\,-\,z_{\sigma}
\partial_l \bar z_{\sigma}\right)^2\right\}
\end{eqnarray}
where $l$ stands for $\tau,x,y$.

The coefficients in front of the terms which break the gauge symmetry $\left(
W_l,Z_l,M_l\right)$ are zero if the order parameter is zero.
The constants $W_l\left(\bar W_l\right)$ are proportional to $t_1$ and
result from the tadpole diagrams with anomalous Green functions, and special
values of the order parameters. The constants
$Z_l$ and $M_l$ are obtained calculating the one-loop diagrams with two
anomalous Green functions. $Z_l$ are proportional to $t^2_1$ and $M_l$ are
proportional to $t^2_2$. One can get the renormalized parameter $g_r$ of the
$CP^1$ model and the renormalized spin-wave velocity $c_r$ calculating the
one-loop diagram with two normal Green functions, and using the Eqs.(44).

Generalized $CP^1$ models with broken gauge symmetry have been largely
discussed in the literature. An additional exitation (thirth
Goldstone boson) appears in the theory as a result of gauge symmetry
breaking. A generalized $CP^1$ model with $"W"$ terms
only has been considered as a model of the spiral phase of a doped
antiferromagnet \cite{16},\cite{17}. 
Due to $"W"$ terms the minimum of the dispersion of the $z_{\sigma},
(\bar z_{\sigma})$ quanta is not at the zero wave-vector, which leads
to incommensurate order.

The massive $CP^1$ model with $"M"$ terms only, has been investigated as a
model of frustrated antiferromagnet, by means of renormalization group
technique and $\frac 1N$ expansion \cite{18},\cite{19}.

As was shown above, the effective model of the frustrated due to the doping
antiferronagnet contains also terms with constants $Z_l$. These terms
split the spectrum of the antiferromafnetic magnons and make the application
of the large $N$ expansion based on $SU(N)$ group questionable. Moreover,
if one considers a theory without massive terms $(t_2=0)$, then a large
$N$ expansion based on $Sp(2N)$ group is plausible.

The magnon fluctuations influence the fermions strongly, and a mean-field
theory which treats fermions separately seems to be not an adequate
way to solve the model. Nevertheless, the effective theory (52) gives a good
intuition for investigation of the effective model Eq.(38) of doped
antiferromagnets.
\ \\

{\bf IV Effective theory in terms of gauge ivariant fields}

Another option is to parametrize the long distance fluctuations with help
of gauge invariant fields. To do this I introduce two gauge invariant
Fermi fields $c_{\sigma}(\bar c_{\sigma})$
\FL
\begin{eqnarray}
c_1(\tau,\vec x) & = & z_1(\tau,\vec x)\psi^B(\tau,\vec x)\,-\,
\bar z_2(\tau,\vec x)\psi^A(\tau,\vec x) \nonumber \\
c_2(\tau,\vec x) & = & \bar z_1(\tau,\vec x)\psi^A(\tau,\vec x)\,+\,
z_2(\tau,\vec x)\psi^B(\tau,\vec x).
\end{eqnarray}

Under the action of the group of rotations the fields $\psi^A(\tau,\vec x),
\psi^B(\tau,\vec x)$ are singlets and the Bose fields $z_{\sigma}(\tau,\vec x)
(\bar z_{\sigma}(\tau,\vec x))$ are spin $\frac 12$ spinors. One can check
that the Fermi fields $c_{\sigma}(\tau,\vec x)\,(\bar c_{\sigma}(\tau,\vec x))$
transform like components of spin $\frac 12$ spinor. Then, it is not difficult
to guess the inverse relations, because there are only two singlets which can
be built up by means of the Fermi $c_{\sigma}(\tau,\vec x)\,,\,
\bar c_{\sigma}(\tau,\vec x)$ and Bose $z_{\sigma}(\tau,\vec x)\,,\,
\bar z_{\sigma}(\tau,\vec x)$ spinors
\FL
\begin{eqnarray}
\psi^B_1(\tau,\vec x) & = & \bar z_1(\tau,\vec x)\,c_1(\tau,\vec x)\,+\,
\bar z_2(\tau,\vec x)\,c_2(\tau,\vec x) \nonumber \\
\psi^A(\tau,\vec x) & = & z_1(\tau,\vec x)\,c_2(\tau,\vec x)\,-\,
z_2(\tau,\vec x)\,c_1(\tau,\vec x).
\end{eqnarray}

The equations (53) can be regarded as a $SU(2)$ transformation; $c_{\sigma}\,
=\,U_{\sigma,\sigma'}\,\psi_{\sigma'},\,\,(\psi_1\,=\,\psi^B,\,
\psi_2\,=\,\psi^A)$ where $U_{11}\,=\,z_1; U_{12}\,=\,-\bar z_2 ; U_{21}\,=\,
z_2 ; U_{22}\,=\,\bar z_1$. Then, it follows that the Fermi measure is
invariant under the change of variables (53) and that the following equalities
hold
\FL
\begin{eqnarray}
& & \bar \psi^A(\tau,\vec x)\psi^A(\tau,\vec x)\,+\,
\bar \psi^B(\tau,\vec x)\psi^B(\tau,\vec x)\,=\,\bar c_{\sigma}(\tau,\vec x)
c_{\sigma}(\tau,\vec x) \nonumber \\
& & \bar \psi^A(\tau,\vec x)\psi^A(\tau,\vec x)
\bar \psi^B(\tau,\vec x)\psi^B(\tau,\vec x)\,=\,
\bar c_1(\tau,\vec x)c_1(\tau,\vec x)\bar c_2(\tau,\vec x)c_2(\tau,\vec x)
\end{eqnarray}

To get the effective action in terms of the fields $c_{\sigma}(\bar c_{\sigma})$
and the unit vector $\vec m\,=\,\bar z\vec\sigma z$, one has to use the relations
\FL
\begin{eqnarray}
\bar c_{\sigma}\partial_{\tau}c_{\sigma} & = & \bar\psi^A D^{(A)}_{\tau}
\psi^A\,+\,\bar\psi^B D^{(B)}_{\tau}\psi^B \nonumber \\
& + & \,\bar\psi^A\psi^B\left(z_1\partial_{\tau} z_2\,-\,z_2\partial_{\tau} z_1
\right)\,-\,\bar\psi^B\psi^A\left(\bar z_1\partial_{\tau}{\bar z}_2\,-\,
\bar z_2\partial_{\tau}\bar z_1\right)
\end{eqnarray}
\FL
\begin{eqnarray}
\frac {1}{2i}\bar c\vec\sigma\,c\cdot\left(\vec m\times\partial_{\tau}
\vec m\right)
\,=\,\bar\psi^A\psi^B\left(z_1\partial_{\tau} z_2\,-\,z_2\partial_{\tau} z_1
\right)\,-\,\bar\psi^B\psi^A\left(\bar z_1\partial_{\tau}{\bar z}_2\,-\,
\bar z_2\partial_{\tau}\bar z_1\right)
\end{eqnarray}
\FL
\begin{eqnarray}
\partial_{\mu}\bar c_{\sigma}\partial_{\mu}c_{\sigma} & = &
\overline{D^{(A)}_{\mu}\psi^A}
D^{(A)}_{\mu}\psi^A\,+\,\overline{D^{(B)}_{\mu}\psi^B}D^{(B)}_{\mu}\psi^B
\,+\,\overline{D_{\mu}z}_{\sigma}D_{\mu}z_{\sigma}\left(\bar\psi^A\psi^A\,+\,
\bar\psi^B\psi^B\right) \nonumber \\
& + & \,\left(\bar\psi^B\partial_{\mu}\psi^A\,-\,\partial_{\mu}
\bar\psi^B\psi^A\right)
\left(\bar z_1\partial_{\mu}\bar z_2\,-\,\bar z_2\partial_{\mu}\bar z_1\right)
\nonumber \\
& - & \left(\bar\psi^A\partial_{\mu}\psi^B\,-\,\partial_{\mu}
\bar\psi^A\psi^B\right)
\left(z_1\partial_{\mu}z_2\,-\,z_2\partial_{\mu}z_1\right)
\end{eqnarray}
\FL
\begin{eqnarray}
\left(\bar c\vec\sigma\partial_{\mu}c\right.  & - &\left. \partial_{\mu}\bar c\vec\sigma
c\right)\cdot\left(\vec m\times\partial_{\mu}\vec m\right)\,=\,
-\,4i \overline{D_{\mu}z}_{\sigma}D_{\mu}z_{\sigma}\left(\bar\psi^A\psi^A\,+\,
\bar\psi^B\psi^B\right) \nonumber \\
& - & \,2i\left(\bar\psi^B\partial_{\mu}\psi^A\,-\,\partial_{\mu}
\bar\psi^B\psi^A\right)
\left(\bar z_1\partial_{\mu}\bar z_2\,-\,\bar z_2\partial_{\mu}\bar z_1\right)
\nonumber \\
& + & \,2i\left(\bar\psi^A\partial_{\mu}\psi^B\,-\,\partial_{\mu}
\bar\psi^A\psi^B\right)
\left(z_1\partial_{\mu}z_2\,-\,z_2\partial_{\mu}z_1\right)
\end{eqnarray}
where $\mu$ stands for $x$ or $y$. Taking into account the above equalities
and Eqs.(44) one obtains
\FL
\begin{eqnarray}
S_{eff}\,=\,\int d\tau d^2x & & \left\{\frac {1}{2g^2}\left(\,\partial_{\tau}
\vec m\cdot\partial_{\tau}\vec m\,+\,c^2\,\partial_{\mu}\vec m
\cdot\partial_{\mu}\vec m\,\right)\,
+\,\bar c_{\sigma}\,\partial_{\tau}\,c_{\sigma}\,+\,
\frac {1}{2m}\partial_{\mu}\bar c_{\sigma}\,\partial_{\mu}c_{\sigma}
\right. \nonumber \\
\,& & \left.+\,i\,\gamma_{\tau}\,\bar c\vec\sigma\,c\cdot\left(\,\vec m\times
\partial_{\tau}\vec m\,\right)\,
 -\,i\,\gamma_r\,\left(\bar c\vec\sigma\,\partial_{\mu}c\,-\,
\partial_{\mu}\bar c\vec\sigma\, c\right)\cdot\left(\,\vec m\times
\partial_{\mu}\vec m\,\right)\right. \nonumber \\
\,& & \left. +\,\frac 12\lambda_0\,
\bar c_1\,c_1\,\bar c_2\,c_2\,+\,\frac {1}{2g^2_0}\,\left(\partial_{\tau}\vec m
\cdot\partial_{\tau}\vec m\,+\,c^2_0\,\partial_{\mu}\vec m
\cdot\partial_{\mu}\vec m\right)\,\bar c_{\sigma}\,c_{\sigma}
\,+\,\mu\,\bar c_{\sigma} c_{\sigma}\right\}
\end{eqnarray}
where
\FL
\begin{eqnarray}
& & g\,=\,2a\sqrt J;\,\,\,\,\,\,\,c\,=\,aJ,\,\,\,\,g_0\,=\,2\sqrt J;\,\,\,\,\,\,\,
c_0\,=\,4a^2J\left(t_1\,+\,2t_2\,-\,\frac {J}{2}\right); \nonumber \\
& & \frac {1}{2m}\,=\,2a^2t_2;\,\,\,\,\,\,\,
\lambda_0\,=\,a^2\left(\frac {4t^2_1}{J}
\,-\,2J\right);\,\,\,\,\,\,\,
\gamma_{\tau}\,=\,\frac {t_1}{J}\,+\,\frac 12;\,\,\,\,\,\,\,
\gamma_r\,=\,a^2\left(\frac {t_1}{2}\,+\,t_2\right).
\end{eqnarray}

What the effective theory Eq.(60) tells us is that, the resulting Fermi
quasiparticles $c_{\sigma}\left(\bar c_{\sigma}\right)$ of the $t_1-t_2-J$
model have both charge and spin. Let us trace the origin of the result. In the
presence of the next to nearest neighbor hopping the dispersion of the charge
carriers (holons) has a two-fold degenerate minimum. One can introduce two
sublattices, and then the charged spinless particles are two, $\psi^A$ and
$\psi^B$. An unexpected result is that in the long-wavelength, low-frequency
limit, these fields can be mapped Eq.(53,54) onto spin $\frac 12$ spinor with
the same charge.

Without the four-fermion term the effective action coincides with the effective
action proposed in \cite{11}. The special point is that the effective model
in \cite{11} is obtained from spin-fermion one which results from a strong-
coupling expansion of the three band $Cu-O$ model. So that, the Fermi
quasiparticles have a transparent physical interpretation, which is not so in
the case of model (60).
\ \\

{\bf V Conclusions}

In this paper a long-wavelength, low-frequency effective theory of $t_1-t_2-J$
model was explicitly derived. The effective action was written as a
generalized $CP^1$ model Eq.(38,42) in terms of bose spinor fields and two
spinless Fermi fields, and as a generalized $\sigma-$model Eq.(60), in terms
of spin one-half Fermi spinor and unit vector.

The $CP^1$ representation seems to be preferable in the light of numerous
theoretical predictions on the nature of disordered ground states in quantum
spin systems. There are reasons to believe that the low energy excitations in
the disordered phase of quantum spin systems are spin one-half deconfined
spinons \cite{21}.

On the other hand, the succesful application of the $\sigma-$model to spin
systems is a reson to favour the generalized $\sigma-$model Eq.(60). The
$\sigma-$model part of the action can be treated in the same way as in
\cite{14}. To deal with the four-fermion term by means of the renormalization
group one has to use techniques described in \cite{21}. The nontrivial point
is the current-current interaction which strongly influences both the spinon
spectrum and the long-wavelength behaviour of the fermions.
\ \\

\acknowledgements

I would like to thank A.Muramatsu and C.K\"ubert for useful discussion
in the course of the work. The hospitality of the Stuttgart University,
and the financial support from Deutscher Akademischer Austauschdienst
are acknowledged.

\end{document}